\documentstyle[12pt]{article}
\begin{document}
\begin{center}

{\bf Three Dimensional Gauge Theory with Topological and Non-topological Mass:
Hamiltonian and Lagrangian Analysis}\\
\vskip 2cm
Subir Ghosh
\footnote {Email address: <subir@boson.bose.res.in>}\\
Physics Department,\\
Dinabandhu Andrews College, Calcutta 700084,\\
India.\\
\end{center}
\vskip 2cm
{\bf Abstract:}\\

Three dimensional (abelian) gauged massive Thirring model is bosonized
in the large fermion mass limit. A further integration of the gauge field
results in a non-local theory. A truncated version of that is the
Maxwell Chern Simons (MCS) theory with a conventional mass term or MCS Proca
theory. This gauge invariant theory
is completely solved in the Hamiltonian and Lagrangian
formalism, with the spectra of the modes determined. Since the vector
field constituting the model is identified (via bosonization) to
the fermion current, the charge current algebra, including the Schwinger term
is also computed in the MCS Proca model.
\newpage
{\bf I. Introduction}
\vskip .5cm

It has been appreciated for quite sometime, that gauge symmetry in
2+1-dimensions is subtle, mainly due the Chern Simons term \cite{d1}.
Self dual theory with the (non-topological) mass term is gauge invariant,
being dual to the Maxwell Chern Simons (MCS) gauge theory. Chern Simons
term is referred to as the topological mass term. A master lagrangian
has been constructed \cite{d2} which can generate both the above mentioned
models.

Including a non-topological mass term in the MCS model leads to the so called
MCS Proca (MCSP) model \cite{k}. This term does not break the gauge invariance
since one can view the MCSP theory as a combination of self dual and Maxwell
theory. The former is equivalent to a gauge theory and the latter is
manifestly a gauge theory. A Lagrangian analysis was given in \cite{k}
where the spectra of two massive modes were provided. In this paper, a
detailed Hamiltonian
constraint analysis \cite{di} is provided for the first time. It is shown
that an involved analysis leads to identical spectra and equations of
motion obtained via Lagrangian method. This is one of our main results.
But there are additional benifits
of Dirac analysis, which we elaborate below.

Let us now put the MCSP model, studied here, in its proper perspective.
Our motivation in the above model is that it has been derived from
a three dimensional $U(1)$ gauged massive Thirring model
\cite{s1} via bosonization
of the fermion modes, (in the large fermion mass limit) \cite{d1,ko}.
The bosonic theory is a master Lagrangian, comprising of the $U(1)$ gauge
field $A_\mu $, and an auxiliary field $B_\mu $, introduced to linearise
the Thirring self interaction term. Integrating over $B_\mu $ leads to
a generalized MCS model, which under certain approximations sheds light
on the self interaction effects on inter-"quark" potential \cite{s2}.
On the other hand, integration of the gauge field $A_\mu $ (in Lorentz
gauge) yields a generalization of the MCSP model in $B_\mu $. A
truncated version of it is the MCSP model in question. The added bonus
of this scheme is that $B_\mu $ {\it reflects the behaviour of the
fermion current $J_\mu=\bar\psi\gamma_\mu\psi $ since $J_\mu\equiv B_\mu /g$,
$g$ denoting the Thirring coupling}. Indeed, we have correctly reproduced
the {\it current conservation} and {\it current algebra} including
the Schwinger term. More complicated fermionic composite objects can also
be studied. This is the other important result of this paper.

The paper is organised as follows: Section II briefly gives the bosonization
results of gauged Thirring model. Section III deals with the Lagrangian
formulation, in a way similar to \cite {k}. The particle spectra is obtained.
Section IV is the main body of our work, consisting of the full Hamiltonian
analysis and current algebra results. The paper ends with a brief conclusion
in Section V.
\vskip 1cm
{\bf II. Bosonization of gauged Thirring model}
\vskip .5cm
The $U(1)$ gauged Thirring model Lagrangian is,
\begin{equation}
{\cal L}_F=\bar\psi i\gamma^\mu (\partial_\mu-ieA_\mu)\psi -m\bar\psi\psi
+{g\over 2}\mid\bar\psi\gamma^\mu\psi\mid^2 -{{pe^2}\over 4}
\mid A_{\mu\nu}\mid^2
+{{qe^2}\over 2}\epsilon_{\mu\nu\lambda}A^\mu A^{\nu\lambda}.
\label{eqlf}
\end{equation}
Here $A_{\mu\nu}=\partial_\mu A_\nu-\partial_\nu A_\mu $ and conventionally one takes $p=1/e^2$,
$q=\mu/(2e^2)$. The reason we have considered them arbitrary will become
clear as we proceed.
The above model is linearised via the auxiliary field $B_\mu$ as
\begin{equation}
{\cal L}_F=\bar\psi i\gamma^\mu (\partial_\mu-ieA_\mu -iB_\mu)\psi
-{1\over{2g}}\mid B_\mu \mid^2 -m\bar\psi\psi
-{{pe^2}\over 4}
\mid A_{\mu\nu}\mid^2 +{{qe^2}\over 2}\epsilon_{\mu\nu\lambda}A^\mu A^{\nu\lambda}.
\label{eqlfb}
\end{equation}
One loop bosonization in the large fermion mass limit
yields the bosonic Lagrangian (to $O(1/m)$),
\begin{equation}
{\cal L}_B=-{a\over 4}C_{\mu\nu}C^{\mu\nu}
+{{\alpha}\over 2}
\epsilon_{\mu\nu\lambda}C^\mu C^{\nu\lambda}-{1\over{2g}}
B_\mu B^\mu
-{{pe^2}\over 4} A_{\mu\nu}A^{\mu\nu}
+{{qe^2}\over 2}
\epsilon_{\mu\nu\lambda}A^\mu A^{\nu\lambda}.
\label{eqlb}
\end{equation}
where  $C_\mu =B_\mu +eA_\mu $, $\alpha =1/(8\pi)$ and $a=-1/(6\pi m)$.
The $U(1)$ gauge invariance present in (\ref{eqlf}) is clearly visible as
regards the $A_\mu $ field. The $A_\mu$	 (gauge) and $B_\mu $ ("matter")
field equations are,
\begin{equation}
a\partial_\mu C^{\mu\alpha}+{{\alpha }\over 2}\epsilon^{\alpha\mu\nu}
C_{\mu\nu}+ep\partial_\mu A^{\mu\alpha}+{{eq}\over 2}\epsilon^{\alpha
\mu\nu}A_{\mu\nu}=0.
\label{eqa}
\end{equation}
\begin{equation}
a\partial_\mu C^{\mu\alpha}+{{\alpha }\over 2}\epsilon^{\alpha\mu\nu}
C_{\mu\nu} -{1\over g}B^\alpha=0.
\label{eqb}
\end{equation}
The above two equations are combined to give
\begin{equation}
{1\over g}B^\alpha+
ep\partial_\mu A^{\mu\alpha}+{{eq }\over 2}\epsilon^{\alpha\mu\nu}
A_{\mu\nu}=0.
\label{eqab}
\end{equation}
Notice that without the gauge field kinetic terms in the parent fermion model,
we would have obtained simply $B_\mu =0$.

The Lagrangian in (\ref{eqlb}) is our
 {\it master} Lagrangian \cite{s1}. Upon selective integration of one of
the interacting fields in turn, different equivalent (dual) theories as
reproduced which are apparantly distinct. In this way, it is possible
to connect different well known theories. The duality between them appears
in the form of particle spectrum, symmetry, Green's function e.t.c..
The next task is to integrate out the gauge field.

\vskip 1cm
{\bf III. Particle spectrum: Lagrangian framework}
\vskip .5cm
Modulo total derivative terms, $A_\mu $ integration in Lorentz gauge gives
\cite {s1},
$$
{\cal L}_B(B_\mu)=B_\mu{{{{ap(a+p)}\over 8}\partial^2+{1\over 2}(p\alpha^2+q^2a)}
\over{({{p+a}\over 2})^2\partial^2+(q+\alpha )^2}}(g^{\mu\nu}
\partial^2-\partial^\mu
\partial^\mu)B_\nu $$
\begin{equation}
+B_\mu
{{{{(p^2\alpha+qa^2)}\over 4}\partial^2+q\alpha (\alpha+q)}
\over{({{p+a}\over 2})^2\partial^2+(q+\alpha )^2}}\epsilon^{\mu\nu\lambda}
B_{\nu\lambda} -{1\over {2g}}B_\mu B^\mu.
\label{eqlb1}
\end{equation}
The equation of motion for $B_\mu $ is
$$
{{{{ap(a+p)}\over 8}\partial^2+{1\over 2}(p\alpha^2+q^2a)}
\over{({{p+a}\over 2})^2\partial^2+(q+\alpha )^2}}(g^{\mu\nu}
\partial^2-\partial^\mu
\partial^\mu)B_\nu $$
\begin{equation}
+
{{{{(p^2\alpha+qa^2)}\over 4}\partial^2+q\alpha (\alpha+q)}
\over{({{p+a}\over 2})^2\partial^2+(q+\alpha )^2}}\epsilon^{\mu\nu\lambda}
B_{\nu\lambda} -{1\over {2g}}B^\mu.
\label{eqllb}
\end{equation}
Clearly $B_\mu $ obeys the current conservation, $(\partial_\mu B^\mu =0)$,
as is required of the fermion current. Defining the dual of $B_\mu $ as
$$^*B_\mu={1\over 2}\epsilon_{\mu\nu\lambda}B^{\nu\lambda},$$
we obtain two equations and the Lagrangian in (\ref{eqlb})
in compact notation as,
$$
{\cal A}(^*B^\alpha)-{\cal D}\partial^2B^\alpha =0,~~~
{\cal A}(B^\alpha)+{\cal D}(^*B^\alpha) =0,$$
\begin{equation}
{\cal L}_B(B_\mu)=B_\mu{\cal A}B^\mu+B_\mu{\cal D}\epsilon^{\mu\nu\lambda}
\partial_\nu B_\lambda.
\label{eqcal}
\end{equation}
The non-local operators are,
\begin{equation}
{\cal A}\equiv
{{{{ap(a+p)}\over 8}\partial^2+{1\over 2}(p\alpha^2+q^2a)}
\over{({{p+a}\over 2})^2\partial^2+(q+\alpha )^2}}\partial^2-{1\over{2g}},
\label{eqa2}
\end{equation}
\begin{equation}
{\cal D}\equiv
2{{{{(p^2\alpha+qa^2)}\over 4}\partial^2+q\alpha (\alpha+q)}
\over{({{p+a}\over 2})^2\partial^2+(q+\alpha )^2}}.
\label{eqd}
\end{equation}
Combining the above equations in (\ref{eqcal}), we get
\begin{equation}
{(\cal A}^2+{\cal D}^2\partial^2)B^\alpha =0.
\label{eqsp}
\end{equation}
Unfortunately the complicated nature of the operators prohibit
further study of
the field equation. Let us now consider the approximations we mentioned
before.

Keeping within the approximations involved in the bosonization scheme
itself we drop $O(a^2)$ terms. However, with a nonvanishing $p$, (that is
in presence of the Maxwell term in (\ref{eqlf})), the non-local nature
of the effective theory persists. In the present case we avoid this
problem by putting $p=0$ and keep only the Chern Simons term in (\ref{eqlf}).
The operators now become,
\begin{equation}
{\cal A}\approx ({{q^2a}\over{2(q+\alpha )^2}}\partial^2-{1\over{2g}}),~~~
{\cal D}\approx {{2q\alpha}\over{q+\alpha}}.
\label{eqad}
\end{equation}
Hence the $B_\mu $ equation reduces to

\begin{equation}
[ ({{q^2a}\over{2(q+\alpha )^2}}\partial^2-{1\over{2g}})^2+
({{2q\alpha}\over{q+\alpha}})^2\partial^2]B^\alpha=0.
\label{eqbp}
\end{equation}
The above equation is "factorised" in the following form \cite{k},
\begin{equation}
(\partial^2+M_+^2)(\partial^2+M_-^2)B^\alpha =0.
\label{eqm}
\end{equation}
The two values of the effective mass parameter are
\begin{equation}
M_{\pm}^2={{(q+\alpha)^2}\over{q^2a}}[{{8\alpha^2}\over a}-{1\over g}
\pm4\alpha ({{4\alpha^2}\over{a^2}}-{1\over{ag}})^{1\over 2}].
\label{eqmpm}
\end{equation}
Substituting the local expressions for ${\cal A}$ and ${\cal D}$, we
arrive at the MCSP model by neglecting
$O(a^2)$ terms but in the above analysis we have not dropped $O(a^2)$
terms. There is no contradiction here since now we are
studying the MCSP model as such, forgetting how it was originated in
the first place. However, is we persist with $a^2\approx 0$
in (\ref{eqbp}), we end up
with a single massive mode,
$$(\partial^2+M^2)B^\alpha =0, ~~M^2\approx {{(q+\alpha)^2}
\over{16q^2\alpha^2g^2}}(1+{a\over{8\alpha^2g}}).$$
This concludes the Lagrangian analysis of the MCSP model.
\vskip 1cm
{\bf IV. Particle spectrum: Hamiltonian framework}
\vskip .5cm
We start with the MCSP lagrangian, using (\ref{eqad}),
\begin{equation}
{\cal L}=PB_{\mu\nu}B^{\mu\nu}+Q\epsilon_{\mu\nu\lambda}B^\mu\partial^\nu B^\lambda
+RB_\mu B^\mu.
\label{eqlh}
\end{equation}
where,
$$
P={{q^2a}\over{4(q+\alpha)^2}},~~~Q={{2q\alpha}\over{q+\alpha}},~~~
R=-{1\over{2g}}.$$
The conjugate momenta and the canonical Hamiltonian are defined in the
standard way,
\begin{equation}
\Pi_\mu={{\partial L}\over{\partial\dot B^\mu}},~~{\cal H}
=\Pi^\mu \dot B_\mu-{\cal L}.
\label{eqmh}
\end{equation}
Explicit expressions for the above are,
\begin{equation}
\Pi_0=0,~~~\Pi_i=-4P(\partial_iB_0-\dot B^i)-Q\epsilon_{ij}B_j,
\label{eqm1}
\end{equation}
\begin{equation}
{\cal H}=-{1\over{8P}}(\Pi+Q\epsilon_{ij}B_j)^2+B_0(\partial_i\Pi_i-
Q\epsilon_{ij}\partial_i B_j)-PB_{ij}B_{ij}-RB_\mu B^\mu.
\label{eqh}
\end{equation}
We now perform the constraint analysis by obtaining the constraints
and subsequently computing the Dirac brackets. Our
aim is to obtain the equations of motion of the modes and reproduce the
spectra obtained in (\ref{eqmpm}).
The primary constraint is
\begin{equation}
\Psi_1(x)\equiv \Pi_0(x)\approx 0,
\label{eqk}
\end{equation}
and time persistence generates the secondary constraint
\begin{equation}
\Psi_2(x)\equiv \dot\Psi_1(x)=[\Psi_1(x),\int d^2y {\cal H}(y)]
=\partial_i\Pi_i(x)-2RB_0(x)-Q\epsilon_{ij}\partial_iB_j(x)\approx 0.
\label{eqkk}
\end{equation}
These brackets are obtained by using the fundamental Poisson brackets
$$[\Pi^\mu(x),B_\nu(y)]=g^\mu_\nu\delta(x-y).$$
The constraints constitute a second class pair with the non-trivial
algebra,
\begin{equation}
[\Psi_1,\Psi_1]=[\Psi_2,\Psi_2]=0,~~[\Psi_1(x),\Psi_2(y)]=2R\delta(x-y).
\label{eqalg}
\end{equation}
The inverse of the constraint matrix $\Psi_{ij}$, defined by
$\int d^2y C_{ij}(x,y)\Psi_{jk}(y,z)=g_{ik}\delta(x-z)$ has the non zero
element
$$C_{12}(x,y)=-{1\over{2R}}\delta(x-y).$$
This generates the nontrivial Dirac brackets
\begin{equation}
[B_0(x),B_i(y)]={1\over{2R}}\partial_i\delta(x-y),~~~
[B_0(x),\Pi_i(y)]=-{Q\over{2R}}\epsilon_{ij}\partial_j\delta(x-y).
\label{eqdb}
\end{equation}
Rest of the brackets are not altered. From now on we will always use
these Dirac Brackets. {\it The $B_0-B_i$ bracket recovers the
correct Schwinger term in the fermion current algebra},
\begin{equation}
[J_0(x),J_0(y)]=[J_i(x),J_j(y)]=0,~~~[J_0(x),J_i(y)]=-{1\over g}\partial_i
\delta(x-y).
\label{eqjj}
\end{equation}

After strong implementation of the constraints ${\cal H}$
in (\ref{eqh}) simplifies to
\begin{equation}
{\cal H}=-{1\over{8P}}(\Pi_i+Q\epsilon_{ij}B_j)^2-PB_{ij}^2
+R(B_0^2+B_i^2).
\label{eqhh}
\end{equation}
{\it Time derivative of $B_0$ reproduces the current conservation},
\begin{equation}
\dot B_0(x)=[B_0(x), H]=\partial_iB_i(x).
\label{eqcc}
\end{equation}
The above current algebra
and conservation equation are two of our main results.
Note that the Dirac brackets are crucial in recovering them.
We now rederive the
particle spectrum.

First we compute the following time derivatives,
\begin{equation}
\dot B_i={1\over{4P}}(\Pi_i+Q\epsilon_{ij}B_j)+\partial_iB_0,
\label{eqbi}
\end{equation}
\begin{equation}
\dot \Pi_i=-{Q\over{4P}}(\epsilon_{ji}\Pi_j+QB_i)
-4P\partial_jB_{ij}-Q\epsilon_{ij}\partial_jB_0 +2RB_i.
\label{eqpi}
\end{equation}
We take time derivatives of the above equations,
$$\ddot B_i={1\over{4P}}(\dot\Pi_i+Q\epsilon_{ij}\dot B_j)
+\partial_i\dot B_0,$$
$$\ddot \Pi_i=-{Q\over{4P}}(\epsilon_{ji}\dot\Pi_j+Q\dot B_i)
-4P\partial_j\dot B_{ij}-Q\epsilon_{ij}\partial_j\dot B_0 +2R\dot B_i.$$
A long algebra yields the following set of equations,
\begin{equation}
[\partial^2-{1\over{2P}}(R-{{Q^2}\over{4P}})]B_i={Q\over{8P^2}}\epsilon_{ij}
\Pi_j,
\label{s1}
\end{equation}
\begin{equation}
[\partial^2-{1\over{2P}}(R-{{Q^2}\over{4P}})]\Pi_i=-2Q\epsilon_{ij}
\partial_j(\partial_kB_k)+{Q\over{2P}}(2R-{{Q^2}\over{4P}})\epsilon_{ij}B_j
+2Q\epsilon_{ik}\partial_j\partial_jB_k-2Q\partial_i(\epsilon_{jk}
\partial_jB_k).
\label{s2}
\end{equation}
The constraints have been used strongly. The same operator
arising in left hand side of both the above
equations is used once again and we get
\begin{equation}
[\partial^2-{1\over{2P}}(R-{{Q^2}\over{4P}})]^2B_i=
-{{Q^2}\over{16P^3}}(2R-{{Q^2}\over{4P}})]B_i.
\label{eqfin}
\end{equation}
The identical equation appears for $\Pi_i$ as well. This chain of derivatives
diagonalizes the equations of motion. Factorising (\ref{eqfin}), we
obtain the {\it identical} set expressions for $M_\pm $ given in (\ref{eqmpm}).
This concludes the Hamiltonian analysis.

Substituting the known expressions we get the explicit forms of the masses,
\begin{equation}
M_\pm^2={{24\pi e^4m}\over{\mu^2}}({1\over{8\pi}}+{{\mu}\over{2e^2}})^2
[{{3m}\over{4\pi}}+{1\over g}\pm{1\over{2\pi}}({{9m^2}\over 4}+{{6\pi m}\over
g})^{1\over 2}].
\label{eqmf}
\end{equation}
Expanding the square root in powers of $(1/m)$, we get
$$
M_+^2\approx {{48\pi e^4m}\over{\mu^2}}({1\over{8\pi}}+{{\mu}\over{2e^2}})^2
({{3m}\over{4\pi}}+{1\over g}),$$
\begin{equation}
M_-^2\approx {{16\pi^2 e^4}\over{\mu^2 g^2}}({1\over{8\pi}}+{{\mu}\over{2e^2}})^2.
\label{eqmf1}
\end{equation}
Interestingly, $M_+>>M_-$ since $M_-$ is {\it independent} of the fermion
mass $m$, the large parameter. Since spin of the particles is determined
by the sign of the mass \cite{d1,k}, the small value of $M_-$ can lead
to a spinless particle.
\vskip 1cm
{\bf V. Conclusions}

Hamiltonian and Lagrangian of the MCS Proca model is performed, with the
full spectra of modes revealed. The interest in the model lies in the fact
that the model has been derived from the bosonized version of a $U(1)$
gauged massive Thirring model. Since the bosonic vector field and the
fermion current are identified, the bosonized model, and in turn the
MCS Proca model, yields properties of its fermion counterpart. As a
non-trivial application of the above, we have computed correctly
the fermion current algebra, with the Schwinger term, staying in the
MCS Proca model framework. Behaviour of other fermionic composite
objects, constructed from fermion currents,
can also be studied in the MCS Proca model, where the quantum
effects enter via the process of bosonization.

\vskip .5cm
\vskip 1cm
{\it Acknowledgment}: I am grateful to Professor S. Dutta Gupta, Director,
S. N. Bose National Centre for Basic Sciences, Calcutta, for allowing
me to use the Institution facilities.

\end{document}